\documentclass[twocolumn,amsmath, amssymb, 10pt, sort&compress, apjl]{aastex631}

\usepackage{xcolor}
\usepackage{graphicx}
\usepackage{bm}
\usepackage{mathtools}
\usepackage{amsmath}
\usepackage{mathrsfs}
\usepackage{orcidlink}
\usepackage{hyperref}
\hypersetup{
 pdftitle={Kilonovae and Long-duration Gamma-ray Bursts}, 
 pdfauthor={Risti\'c}, 
 colorlinks=true,
 citecolor={blue},
}

\newcommand{\GRBxx}[1]{\mbox{GRB\,#1}}

\graphicspath{{figures/}}

\begin{document}

\title{Kilonovae and Long-duration Gamma-ray Bursts}

\author[0000-0001-7042-4472]{Marko Risti\'c}
\thanks{ISTI Fellow}
\email{mristic@lanl.gov}
\affiliation{Theoretical Division, Los Alamos National Laboratory, Los Alamos, NM 87545, USA}
\affiliation{Center for Theoretical Astrophysics, Los Alamos National Laboratory, Los Alamos, NM 87545, USA}

\author[0000-0002-8825-0893]{Brandon L. Barker}
\altaffiliation{Metropolis Fellow}
\affiliation{Center for Theoretical Astrophysics, Los Alamos National Laboratory, Los Alamos, NM 87545, USA}
\affiliation{Computational Division, Los Alamos National Laboratory, Los Alamos, NM 87545, USA}

\author[0000-0003-1758-8376]{Samuel Cupp}
\affiliation{Theoretical Division, Los Alamos National Laboratory, Los Alamos, NM 87545, USA}
\affiliation{Center for Theoretical Astrophysics, Los Alamos National Laboratory, Los Alamos, NM 87545, USA}

\author[0000-0002-7893-4183]{Axel Gross}
\affiliation{Theoretical Division, Los Alamos National Laboratory, Los Alamos, NM 87545, USA}
\affiliation{Center for Theoretical Astrophysics, Los Alamos National Laboratory, Los Alamos, NM 87545, USA}

\author[0000-0003-1707-7998]{Nicole Lloyd-Ronning}
\affiliation{Center for Theoretical Astrophysics, Los Alamos National Laboratory, Los Alamos, NM 87545, USA}
\affiliation{Computational Division, Los Alamos National Laboratory, Los Alamos, NM 87545, USA}

\author[0000-0003-4156-5342]{Oleg Korobkin}
\affiliation{Theoretical Division, Los Alamos National Laboratory, Los Alamos, NM 87545, USA}
\affiliation{Center for Theoretical Astrophysics, Los Alamos National Laboratory, Los Alamos, NM 87545, USA}

\author[0000-0001-6432-7860]{Jonah M. Miller}
\affiliation{Center for Theoretical Astrophysics, Los Alamos National Laboratory, Los Alamos, NM 87545, USA}
\affiliation{Computational Division, Los Alamos National Laboratory, Los Alamos, NM 87545, USA}
\affiliation{Michigan SPARC, Los Alamos National Laboratory, Ann Arbor, MI 48105, USA}

\author[0000-0002-9950-9688]{Matthew R.\ Mumpower}
\affiliation{Center for Theoretical Astrophysics, Los Alamos National Laboratory, Los Alamos, NM 87545, USA}
\affiliation{Computational Division, Los Alamos National Laboratory, Los Alamos, NM 87545, USA}
\affiliation{Department of Physics and Astronomy, University of Notre Dame, Notre Dame, IN, 46656, USA}
\affiliation{Obsidian Research, Fort Wayne, IN 46835, USA}

\date{\today}

\begin{abstract}
Recent detections of kilonova-like emission following long-duration gamma-ray bursts GRB211211A and GRB230307A have been interpreted as originating from the merger of two neutron stars. 
In this work, we demonstrate that these observations are also consistent with nucleosynthesis originating from a collapsar scenario. 
Our model is capable of reproducing the observed optical and infrared light curves using a single weak $r$-process component. 
The absence of lanthanide-rich material in our model, consistent with the data, challenges the prevailing interpretation that a red evolution in such transients necessarily indicates the presence of heavy $r$-process elements. 
\end{abstract}


\section{Introduction}

For over three decades, it has been known that the gamma ray burst (GRB) duration distribution is bimodally split, with so-called ``short'' GRBs lasting $\lesssim 2$ seconds, and ``long'' GRBs lasting $\gtrsim 2$ seconds \citep{1993ApJ...413L.101K}. Long GRBs have been tied to the collapse of massive star progenitors (collapsars) through, among other arguments, observations of coincident supernova in their light curves and spectra \citep{Gal98,Hjorth03,WB06,HB12} as well as their locations in star forming regions in their host galaxies \citep{BKD02,Fong2010, Fong2013, Ly17}.  The progenitors of short GRBs were postulated to be compact object mergers (double neutron star, or neutron star-black hole) and this was put on firm footing with coincident multimessenger detection of a short-GRB and gravitational waves from a neutron star merger \citep{2017PhRvL.119p1101A, 2017ApJ...848L..13A}.\\

Recently, however, a handful of long GRBs have been categorized as originating from compact object mergers. Two notable examples include \GRBxx{211211A} \citep{Rastinejad2022} and \GRBxx{230307A} \citep{Yang24, Levan24}, with prompt durations lasting about 40 seconds in each case. In determining the progenitors of these GRBs, these studies present the presence of late-time infrared emission as a key signature indicating heavy-element nucleosynthesis stemming from a compact object merger. Additionally, \cite{Levan24} present JWST spectra of GRB230327A at 29 and 61 days, showing evidence of an emission line at 2.15 microns which they interpret as {Te III}. The spectra also suggest the potential presence of {W III} and {Se III} at these later times.  These interpretations suggests heavy element production at the site of a long GRB.\\

Many of the previous discussions of long GRBs with kilonova signatures have explored the viability of a binary neutron star merger progenitor for these events, with nucleosynthesis occurring from the extant neutron-rich environment of a tidally disrupted neutron star \citep{Rastinejad2022, Yang24, Troja22, Yang2022, Zhang2022, Levan24, Sun2025, Gottlieb25}, although see \cite{Barnes2023} and \cite{Yang2022} who show the possibility of a collapsar and white dwarf-neutron star merger origin, respectively, for GRB 211211A. The duration, $\tau$, of the prompt gamma-ray emission reflects the lifetime of the jet in the system\footnote{More precisely, the observed duration is the timescale over which the engine operates after the jet breaks out of the star, as in equation 2 of \cite{Brom12}.}, for cases where the prompt emission arises from internal dissipation processes in the jet. This lifetime has traditionally been estimated as the amount of material available in the disk ($M_{disk}$) divided by the accretion rate ($\dot{M}$) of this material onto the central engine:  $\tau \sim M_{disk}/ \dot{M}$, which puts a limitation on the duration from systems that don't have the material to form a disk massive enough to sustain a jet for long durations. Recent studies have shown \citep{Gottlieb25} that compact object mergers can make GRBs on the order of 10's of seconds, where the duration of the GRB (under assumptions of the jet energy, magnetic flux threading the black hole, the radiative efficiency and jet opening angle) is instead fundamentally set by the timescale of the build-up of magnetic flux up to a magnetically arrested disk (MAD) state. Although very intriguing, we think it is still worth continued consideration of a non-compact object merger progenitor for these long GRBs.
 
 Galactic chemical evolution studies have also shown the presence of heavy element production very early in the history of our galaxy \citep[e.g.][and references therein]{Cote19}.  A recent study by \cite{Sal25} showed that even when allowing for the shortest possible merger delay-time distributions in binary neutron star and neutron star-black hole mergers, these compact binary mergers alone cannot account for the observed heavy metal enrichment in metal poor stars in our galaxy. \\

Additionally and compellingly, \cite{Neg25} and \cite{Zhu25} have shown that when GRB211211A and GRB230307A are analyzed in a detailed high-dimensional parameter space --- accounting for both spectral and fine-scale temporal information over the duration of the burst ---  both of these GRBs appear to align with the phase space populated by GRBs associated with collapsars\footnote{See also \cite{Barnes2023}} \citep[i.e. the region where those GRBs that have an observed coincident supernova fall; see Figure 2 and the Appendices of][]{Neg25}. More recently, \cite{Arunachalam25} have concluded that the blackbody continuum in the GRB230307A 29-day spectrum is either not due to dust emission or GRB230307A did not originate from a binary compact-object merger. \\

The collapsar system \citep{Woos93, MW98} is a collapsing massive star that produces a rapidly rotating black hole-accretion disk central engine that launches a relativistic outflow through the so-called Blandford-Znajek process \citep{BZ77, MT82}. In this process, rotation and frame-dragging effects near the black hole horizon cause a build-up of Poynting flux along the spin axis of the black hole.
The powerful jet can carry enough energy and momentum to punch through the surrounding stellar material and propagate into the circumstellar medium where it is inferred from observations to have Lorentz factors on the order of hundreds to thousands \citep[see, e.g.,][and references therein]{Ghir18}.  
As the jet is launched from the central engine and traverses the in-falling stellar material, a so-called ``cocoon'' region is created - a hot, dense region surrounding the jet, but with smaller outflow velocity \citep{RR02, Brom11, Brom14}. 

\cite{Mum25} have recently proposed a new process and site for the production of heavy elements: the complex interaction of light and matter within a relativistic GRB jet stemming from a collapsar. 
\cite{Mum25} showed that as the jet is crossing the dense stellar material, if the flux of high energy photons is sufficient\footnote{This point was discussed in \cite{Mum25}; there is the important issue of the optical depth to pair production suppressing the high energy tail.  In that paper, they provide a set of arguments that suggest in some cases the high energy photon flux may achieve the necessary level to allow the photo-hadronic nucleosynthesis process to take place. For now, we treat the high energy photon flux as an assumption. This however, is a key assumption, which will be modeled in detail in a future publication.}, photons will interact with the shock compressed material at the jet head.  
At the interface between the photons and the shocked stellar material, high-energy, photohadronic interactions may lead to an excess of neutron production at the jet head on dynamical timescales $\tau \sim \mathcal{O}(10^{-4})$ s.
These neutrons can then propagate into the hot, dense cocoon region, allowing for rapid neutron capture process ($r$-process) nucleosynthesis within the cocoon.

In this paper, we show how this model of nucleosynthesis in this collapsar scenario can reproduce the observations of kilonovae in GRBs 211211A and 230307A.  Our paper is organized as follows: in Section \ref{sec:methods}, we describe the methods we use to compute the predicted $r$-process signatures around the jet head, both under some basic assumptions without prior knowledge of the observations and with a more thorough Bayesian inference analysis.  In Sections \ref{sec:results} and \ref{sec:discussion}, we show how these light curves reproduce the observed data remarkably well and present discussion of the implications of these results. Finally, in Section \ref{sec:conclusion}, we present our conclusions.

\section{Methods}
\label{sec:methods}

\subsection{Nucleosynthesis Calculations}

\begin{figure}[t]
  \centering
  \includegraphics[width=\columnwidth]{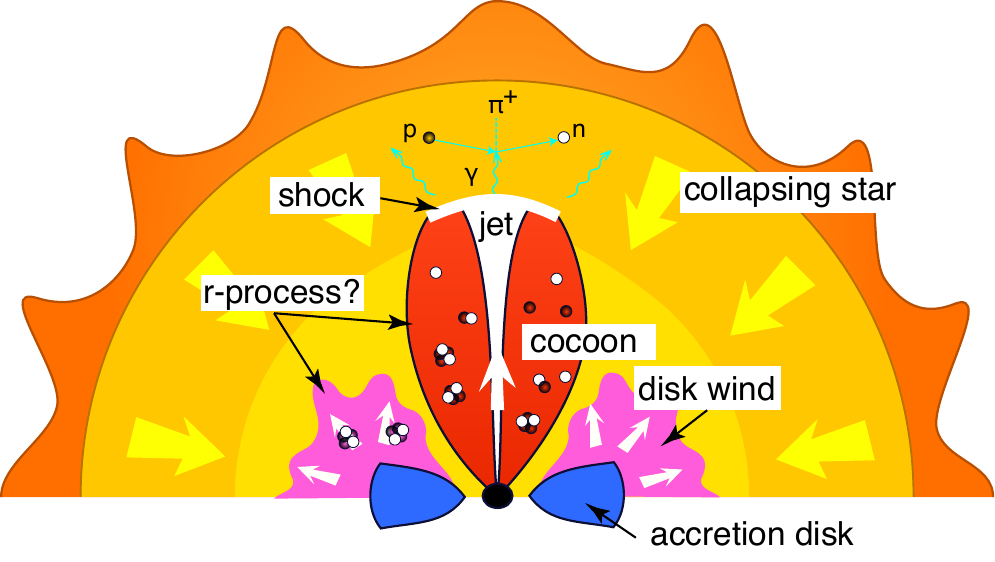}
  \caption{Schematic of a collapsar showing the two proposed sites for neutron-rich nucleosynthesis, emanating from the jet/cocoon or from the disk wind. }
\label{fig:collapsar}
\end{figure}

In the standard picture of collapsars, nucleosynthesis originates in winds that arise after the creation of an accretion disk \citep{MacFadyen1999, Siegel2019,Miller2020,Just2022,Gottlieb2022,Issa2025}. 
This long explored mechanism is capable of producing  both light and heavy $r$-process material depending on the treatment of the relevant neutrino physics \citep{Surman2006}. 
The interaction of matter with high-energy photons represents another avenue for creating neutron-rich conditions associated with the jet \citep{Mum25}. 
The two sites for nucleosynthesis are shown in Figure \ref{fig:collapsar}.

\begin{figure}[t]
  \centering
  \includegraphics[width=\columnwidth]{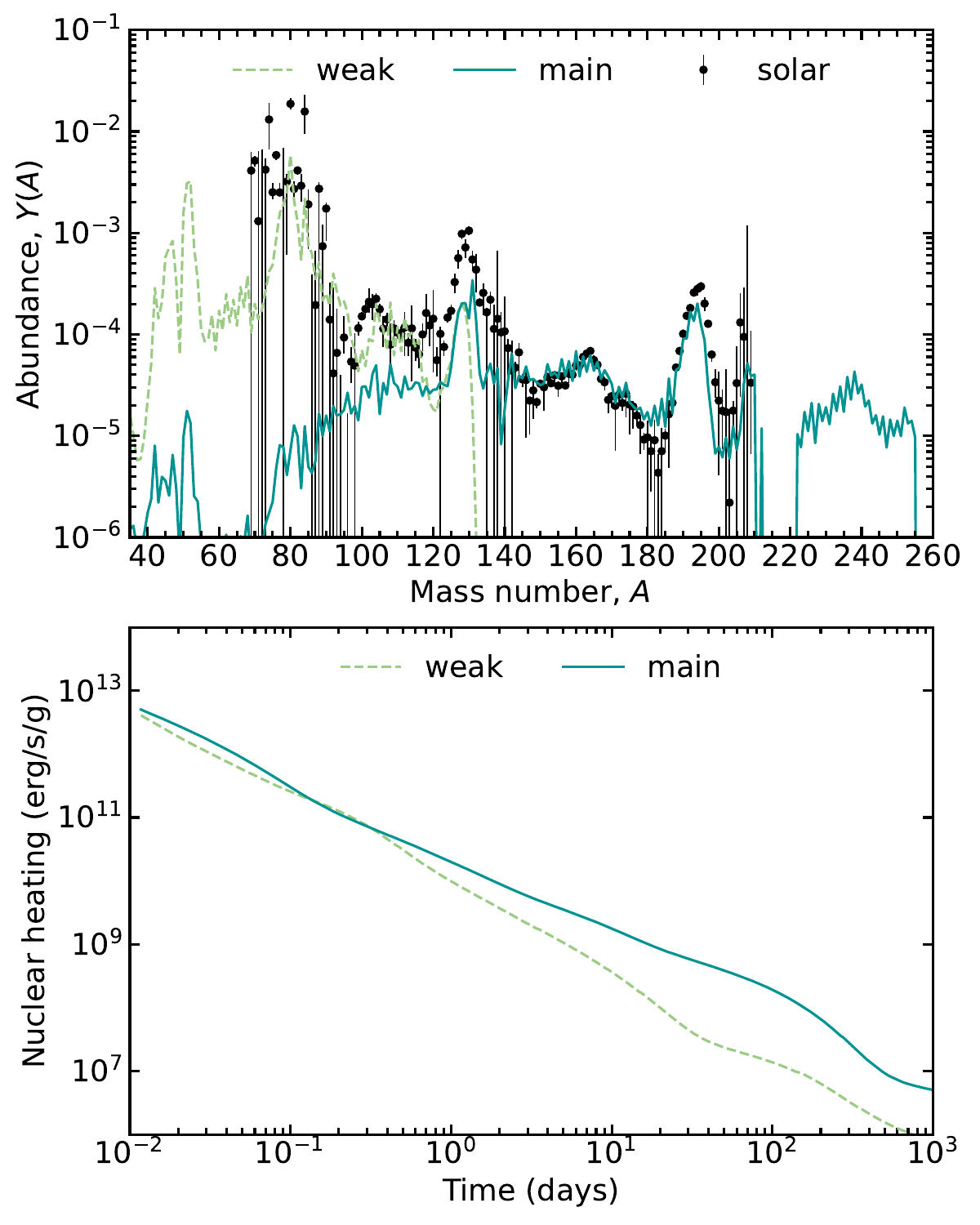}
  \caption{Composition (at one day) and heating in collapsar nucleosynthesis, illustrating both weak and main $r$-processes. }
\label{fig:nucleosyn}
\end{figure}

We perform nucleosynthesis calculations with version 1.6.0 of the Portable Routines for Integrated nucleoSynthesis Modeling (PRISM) reaction network \citep{Sprouse2021}. 
We consider fiducial models for nucleosynthesis from \citep{Mum25}.
These models span two possible outcomes: a weak $r$-process with material up to $A\sim130$, a complete main $r$-process that produces second and third peak elements in addition to actinides. 
The nuclear heating that enters into the kilonova modeling is self-consistently calculated with evolution of abundances as in \cite{Zhu2021}. 
The composition and heating for this work is shown in Figure \ref{fig:nucleosyn}. 

\subsection{Kilonova Model}
\label{sec:kne-model}

Our single-component kilonova model consists of a spherically symmetric ejecta component introduced in \cite{Wollaeger2018}. 
The ejecta is taken to be homologously expanding and is parameterized by mass $m_{\rm{ej}}$ and velocity $v_{\rm{ej}}$. 
The spatial discretization consists of 64 radial cells and a maximum grid velocity twice that of $v_{\rm{ej}}$; Figure~2 in \cite{2021ApJ...910..116K} displays the ``S" schematic corresponding to this morphology. 

We use \texttt{SuperNu} \citep{SuperNu}, a Monte Carlo code for simulation of time-dependent radiation transport with matter in local thermodynamic equilibrium, to generate simulated kilonova spectra. 
The ejecta is assumed to have fixed composition and morphology for the duration of each simulation. 
The radioactive heating contributions are also weighted by thermalization efficiencies introduced in \cite{Barnes_2016} (see \cite{Wollaeger2018} for a detailed description of the adopted nuclear heating). We use detailed opacity calculations via the tabulated, binned opacities generated with the Los Alamos suite of atomic physics codes \citep{2015JPhB...48n4014F,2020MNRAS.493.4143F,nist_lanl_opacity_database}. 

The \texttt{SuperNu} outputs in this work are post-processed to a source distance of $10$ pc, in units of erg s$^{-1}$ cm$^{-2}$ \AA$^{-1}$. The spectra are binned into 1024 equally log-spaced wavelength bins spanning $0.1 \leq \lambda \leq 12.8$~microns. 
For the purposes of comparison with observations, we convolve our simulated \texttt{SuperNu} spectra with the Rubin Observatory $grizy$ ($\sim 4000 - 11000$ \AA) and 2MASS $JHK$ ($\sim 11000 - 24000$ \AA) broadband filters to generate broadband light curves in these filters.

\subsection{Fiducial Simulation Grid}

We generate a grid of \texttt{SuperNu} simulations characterized by the fiducial ejecta parameters described in Table \ref{tbl:fiducial_params}. 
These fiducial parameters span the same ejecta parameter range as in previous kilonova studies of neutron star mergers \citep[e.g.][]{Coughlin18, kilonova-lanl-WollaegerNewGrid2020, Ristic22, Gillanders23, Ristic23, Peng24, Desai25, Koehn25}. 
In comparing directly to the parameters associated with the assumption of a neutron-star-merger kilonova engine, we aim to explore our model's ability to describe the GRB211211A and GRB230307A observations comparably with minimal prior assumptions.

\begin{deluxetable}{cc}[h!]
\tablehead{
\colhead{\hspace{.75cm}Ejecta Parameter}\hspace{.5cm} & \colhead{\hspace{.5cm}Fiducial Values}\hspace{.75cm} }
\startdata
Mass $m_{\rm{ej}}$ & $0.001, 0.01, 0.1$ [$M_\odot$]  \\
Velocity $v_{\rm{ej}}$ & $0.1, 0.2, 0.3$ [$c$] \\
$r$-process Composition & weak, main \\
\enddata
\caption{Fiducial simulation grid ejecta parameters.}
\label{tbl:fiducial_params}
\end{deluxetable}

\subsection{Optimized Extension of Fiducial Grid}
\label{sec:surrogates}

We then build upon the grid of fiducial parameter simulations presented in Table \ref{tbl:fiducial_params} and expand our initial library of models to one suitable for surrogate model training.
We choose new simulation parameters iteratively by fitting a two-dimensional parabola in $[m_{\rm{ej}}$, $v_{\rm{ej}}]$ space to the log-likelihood ($\ln \mathcal{L}$) associated with each simulation, defined as
\begin{equation}
\label{eq:lnl_parabola}
    \ln \mathcal{L} = \sum_{b, t} -\frac{1}{2} \frac{(y_{b ,t} - \hat{y}_{b, t})^2}{\sigma^2},
\end{equation}
where $y_{b, t}$ corresponds to the observation in a given broadband filter $b$ at time $t$, $\hat{y}_{b, t}$ the simulation prediction for the same $b$ and $t$, and $\sigma$ the observation uncertainty, all in units of AB magnitudes.

We apply the parabolic fit to $\ln \mathcal{L}$ of both GRBs, and iteratively place new simulations in unequally-sized batches based on the fit's prediction of the peak $\ln \mathcal{L}$ for each GRB given the current iteration of the simulation grid.
We expand the dynamic range of our parameters during grid refinement, specifically allowing velocities as low as $v_{\rm{ej}} = 0.01c$ and masses as high as $m_{\rm{ej}} = 0.5 M_\odot$.
The final simulation grid consists of 59 total simulations spanning ejecta parameters $10^{-3} \leq m_{\rm{ej}}/M_\odot \leq 0.5$ and $0.01 \leq v/c \leq 0.3$, sufficient for surrogate model training. 

Given the success of Gaussian processes (GPs) in similar prior studies \citep[e.g.][]{Coughlin18, Ristic22}, we use the \texttt{scikit-learn} \citep{scikit-learn} GP regressor as our surrogate model for this study. 
We instantiate each GP's hyperparameters as follows: GP length scales are determined by the standard deviation of the inputs $(m_{\rm{ej}}, v_{\rm{ej}})$, while the length scale bounds are set by the minima and maxima of the inputs, and our kernel is a product of a \texttt{WhiteKernel} and a radial basis function \texttt{RBF} kernel.
A separate GP is trained for each of the \textit{grizJK} broadband filters considered in this study.
Each GP is trained on 54 of the 59 simulations, with five simulations reserved for the test set to ensure fitting fidelity.

For inference, we use the RIFT \citep{RIFT} parameter inference algorithm and compare to the observations of each GRB originally reported in \cite{Rastinejad2022} and \cite{Yang24}.
Using the adaptive volume sampler in RIFT \citep{AVsampler, Wagner25}, we optimize over the log-likelihood
\begin{equation}
    \ln \mathcal{L} = \sum_{b, t} -\frac{1}{2} \frac{(y_{b ,t} - \hat{y}_{b, t})^2}{\sigma^2 + \sigma_{\rm{sys}}^2},
\end{equation}
where all terms are defined as in Equation \ref{eq:lnl_parabola} with the addition of an inference parameter $\sigma_{\rm{sys}}$ which encapsulates our light-curve fitting systematic uncertainty in units of AB magnitudes. 
During inference, we scale our \texttt{SuperNu} light curves from absolute to apparent AB magnitudes according to the source distances reported in \cite{Rastinejad2022} and \cite{Yang24}, respectively.
We terminate our inference analyses upon accruing $\sim 10^3$ effective samples.

\section{Results}
\label{sec:results}

\subsection{Initial Comparison to Fiducial Light Curves}

\begin{figure*}
    \centering
    \includegraphics[width=0.49\linewidth]{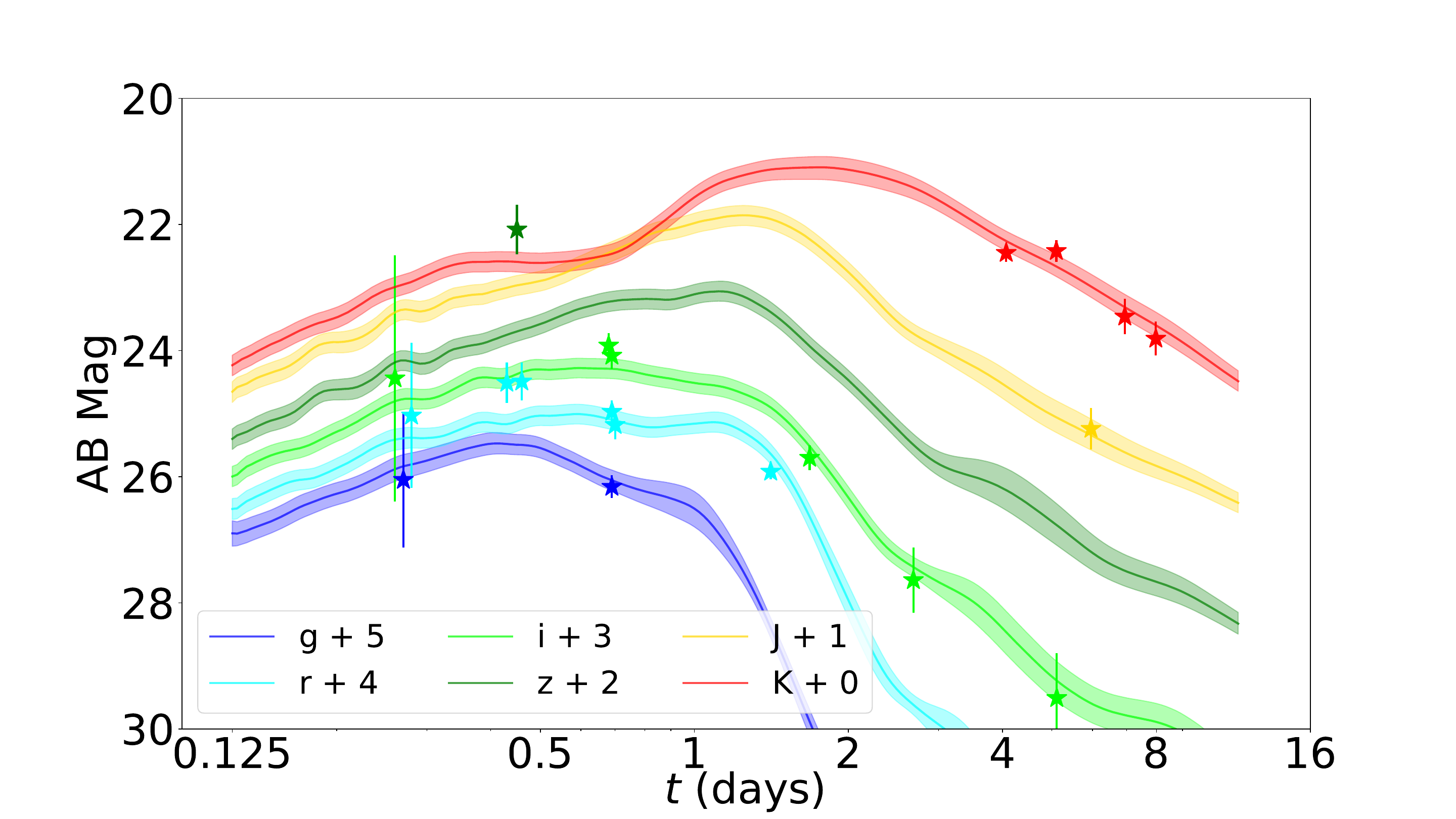}    
    \includegraphics[width=0.49\linewidth]{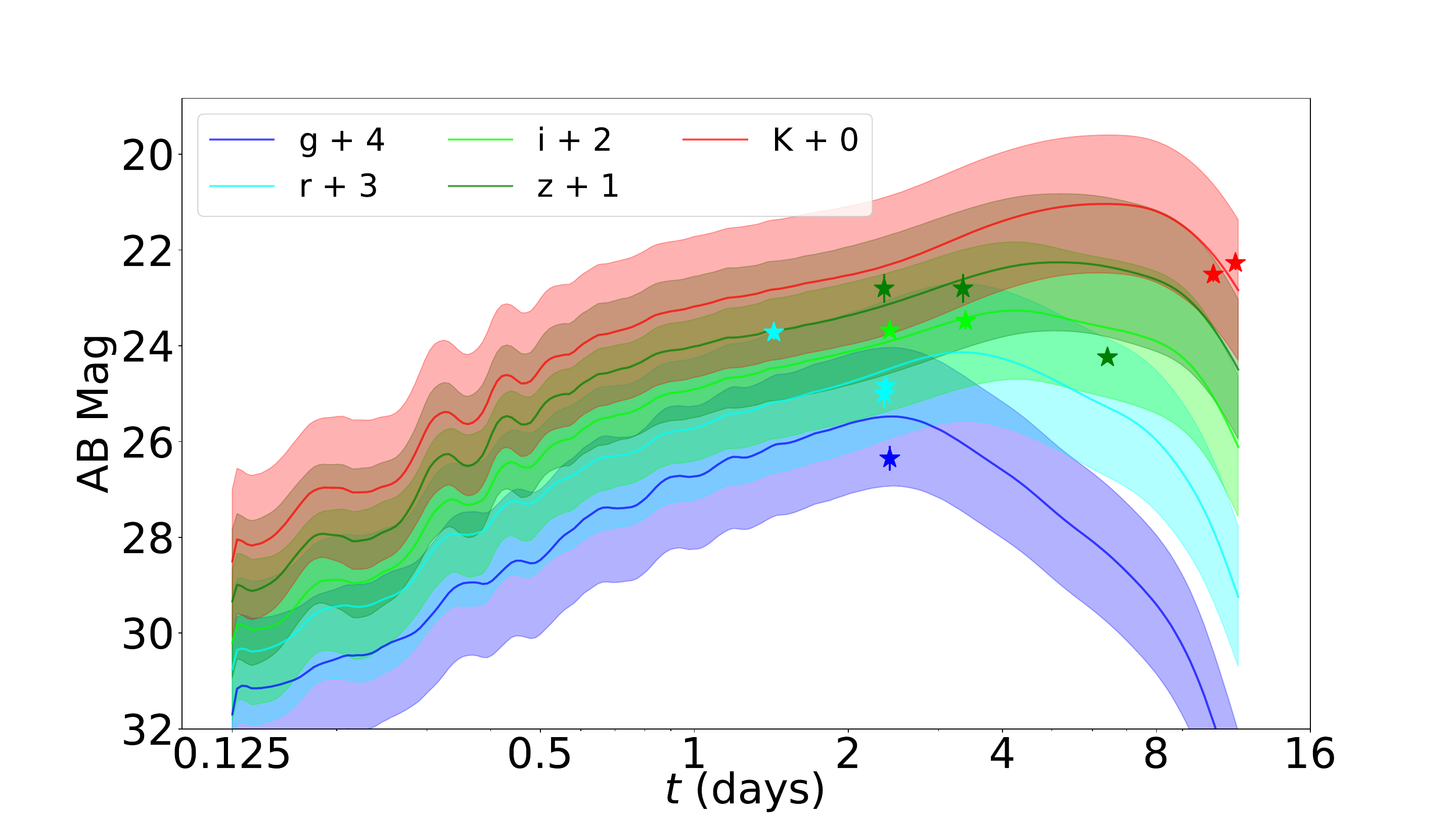}
    \includegraphics[width=0.49\linewidth]{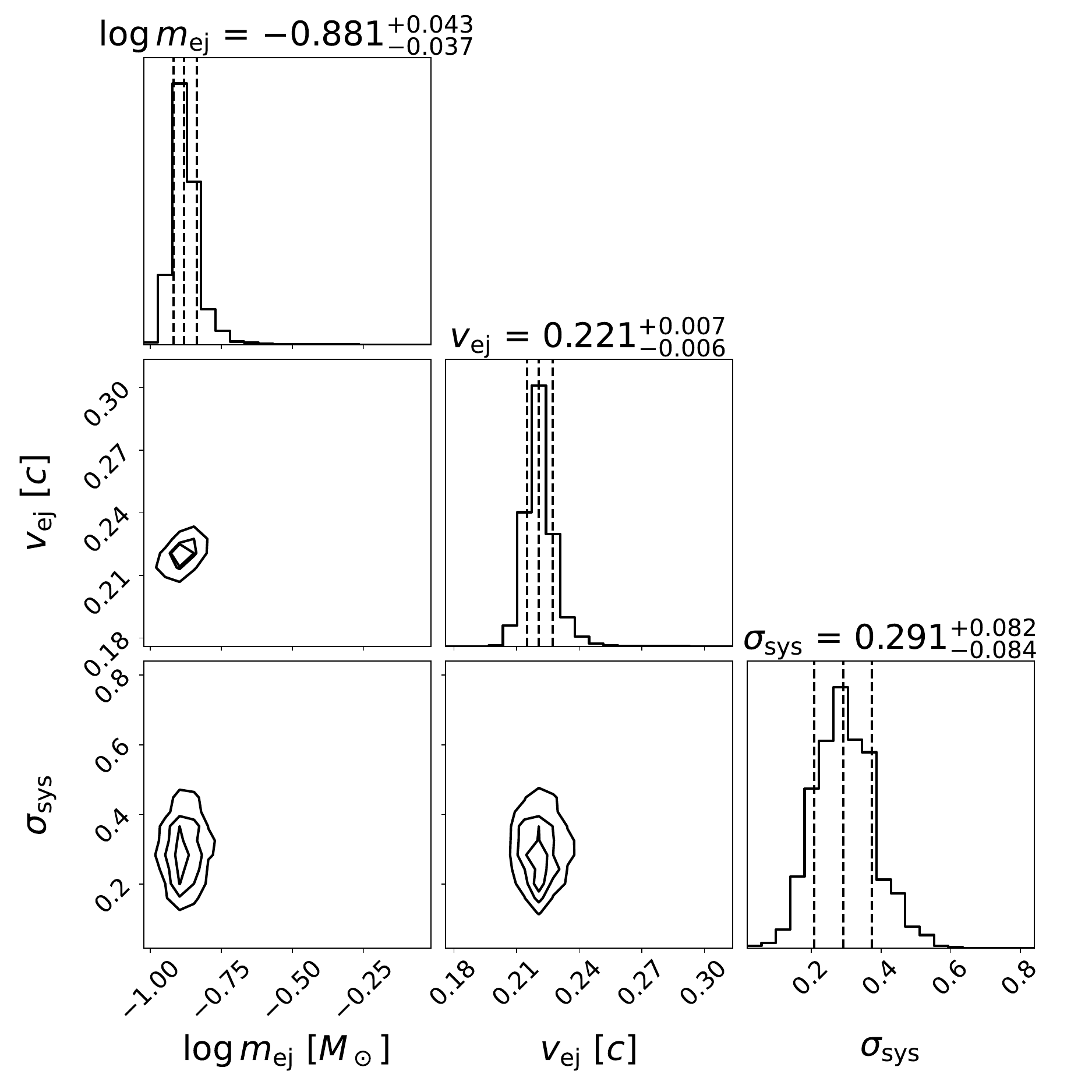}
    \includegraphics[width=0.49\linewidth]{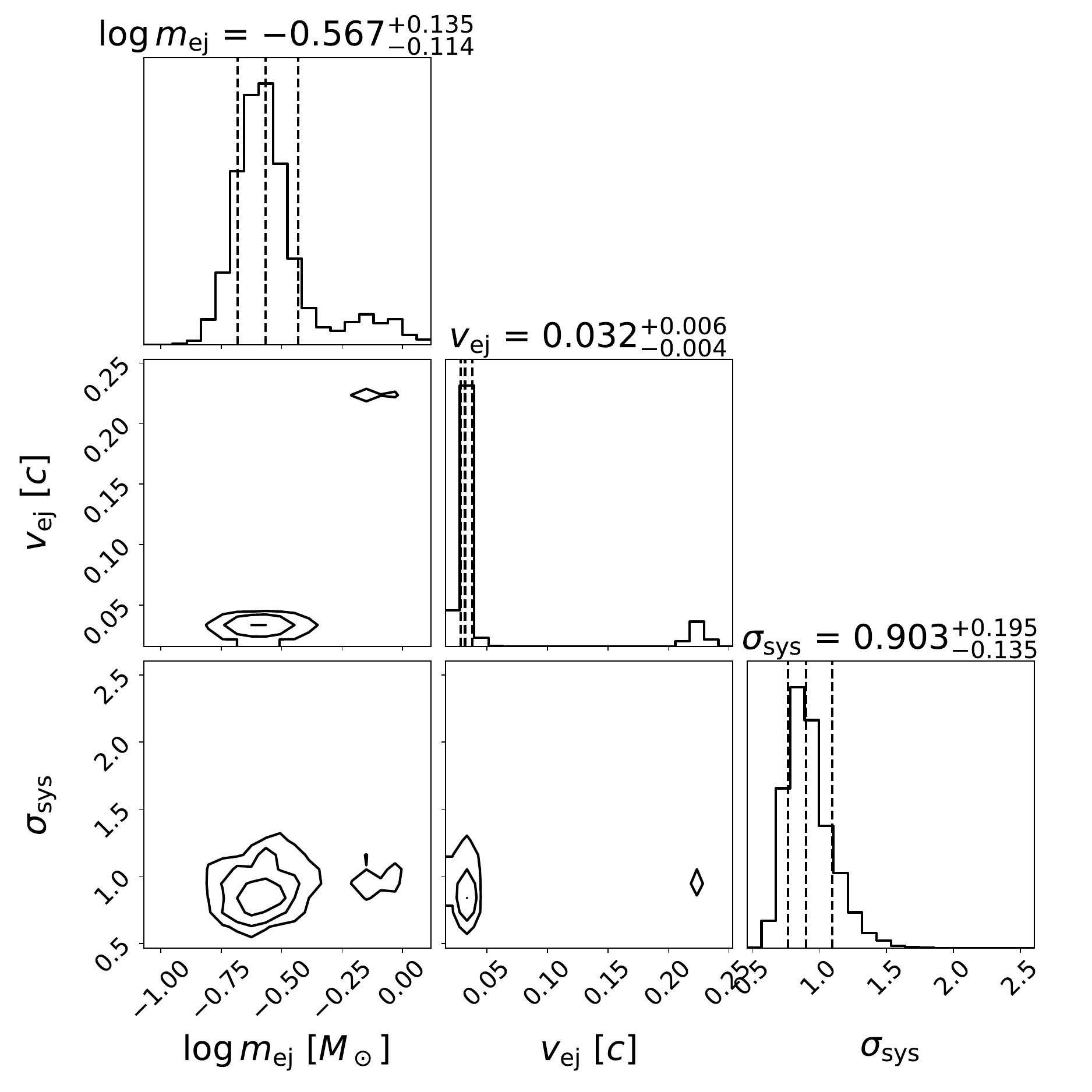}
    \caption{\GRBxx{211211A} (left) and \GRBxx{230307A} (right) optical/near-infrared observations compared to the best-fit interpolated light curve characterized by the parameter values reported in the posterior plots under the assumption of a weak $r$-process ejecta composition. The $r$ and $i$ observations at 0.27 days are artificially separated by $\pm 0.01$ days for visual clarity.}
    \label{fig:lc_inference}
\end{figure*}

We perform an initial comparison of the broadband light curve predictions for our model using fiducial parameters of ejecta outflow (Table~\ref{tbl:fiducial_params}) to the optical and near-infrared observations associated with \GRBxx{211211A} and \GRBxx{23037A}, compiled from \cite{Rastinejad2022} and \cite{Yang24}, respectively,
with subtracted contribution from the afterglow model.
We qualitatively identify that our fiducial model with ejecta outflow parameterized by $m_{\rm{ej}} =  0.1 M_\odot$ and $v_{\rm{ej}} = 0.2c$ consisting of a weak $r$-process composition reproduces the light curves of both GRBs.
Most notably, prior to any data fitting or fine-tuning of our model parameters, we immediately observe that the $m_{\rm{ej}} =  0.1 M_\odot$, $v_{\rm{ej}} = 0.2c$ model light curves match \emph{both} the blue and red emission for both GRBs using only a single ejecta component. 
These ejecta values are consistent with expectations for collapsar ejecta in the literature \citep{Fujimoto07, Barnes2023}.

\subsection{Parameter Inference with Refined Simulation Grid}
\label{sec:inference}

Guided by the fiducial parameter fit, we generate light-curve surrogate models as described in Section \ref{sec:surrogates} and recover Bayesian posterior distributions for $m_{\rm{ej}}$, $v_{\rm{ej}}$, and $\sigma_{\rm{sys}}$ for each of the GRBs in this study.
Figure \ref{fig:lc_inference} presents our best-fit light curves and the associated ejecta and $\sigma_{\rm{sys}}$ parameters which describe them. 
As predicted by the naive fiducial light curve fit described earlier, both sets of observations are recovered well by our model's light curves.
We find that GRB211211A is well represented by a lighter $m_{\rm{ej}} = 0.13 M_\odot$, fast-moving $v_{\rm{ej}} = 0.221c$ ejecta, while GRB230307A favors a more massive $m_{\rm{ej}} = 0.27 M_\odot$, slower $v_{\rm{ej}} = 0.03c$ ejecta, with the latter result corroborated by previous inference of GRB230307A using semi-analytic models \citep{Gillanders23}.
We note that the velocity posteriors for GRB230307A do exhibit weak support for a fast-moving ejecta comparable to that of GRB211211A, with the faster-velocity ($\sim 0.2c$) distributions for both GRBs consistent with the slower end of the velocity distribution found in hydrodynamical simulations of both choked and successful collapsar jets \citep[Figs. 2 \& 3 of][]{Eisenberg22}.
In our study, we also examine the light curves associated with an intermediate and full $r$-process, but find that these compositions are unable to recover the blue emission successfully regardless of ejecta kinematic parameters.

\section{Discussion}
\label{sec:discussion}

The relatively large ejecta masses favored in our fits are, in principle, also possible in a scenario in which nucleosynthesis is occurring in the disk of the collapsar \citep[e.g.][]{Siegel2019}. 
However, the high velocities (GRB211211A) and sharply-peaked velocity posteriors (GRB230307A) predicted in our inference disfavor this scenario, as disk winds are typically slower-peaked with long tails \citep{Miller2020}.
Moreover, these ejecta masses rule out double neutron star mergers as the source of the nucleosynthesis, and are at the very edge of what is reasonably possible for a black hole-neutron star merger progenitor \citep{Fern17,Curtis2023}.

In Section \ref{sec:inference}, we identify that our model is capable of matching the blue and red emission simultaneously using a single ejecta component. 
This match is in contrast to neutron-star-merger kilonovae, which require at least two separate ejecta components with appreciably different parameterizations to successfully fit both early-time blue and late-time red emission \citep[see e.g.,][]{Cowperthwaite17, Evans2017, Kasliwal17, Smartt2017, Tanvir2017, Troja2017, Villar17}.
Moreover, for GRB~211211A, \cite{Hamidani24} could not find a good fit even with a two-component kilonova model, and proposed a scenario where the emission is additionally powered by a late central engine activity.
The blue emission in our model arises from the weak $r$-process composition producing lighter elements than the lanthanides, 
resulting in a low opacity ejecta which permits the propagation of higher energy, blue photons; 
the lack of lanthanides does not impede late-time infrared emission since the ejecta mass is sufficiently high compared to neutron star merger ejecta to allow for the emission of infrared thermal photons.
Similarly, the presence of a near-infrared emission ``bump'' (that exceeds the fading afterglow flux in this energy band)
has previously been used as evidence of the production of lanthanides, indicating the presence of $r$-process nucleosynthesis \citep[e.g.][]{Yang15, Jin15}. 
Recent studies, however, have shown that late-time infrared emission can occur even without the $r$-process. 
For example, \cite{Tak23} and \cite{2024ApJ...961....9F} found that different assumptions for the ejecta velocity distribution can yield late-time features in the light curves which can mimic those typically associated with heavy-element production.
Given that the cocoon may span a wide range of velocities, and that the choice of velocity distribution can have significant observational impact, we note that our single component should be interpreted as a simplified representation of a more complex underlying velocity distribution.
Building upon the hypothesis of a white-dwarf neutron-star (WDNS) merger as the GRB211211A progenitor \citep{Yang2022}, 
\cite{LiuXX25} find that a WDNS merger produces nucleosynthetic yields encompassing only a fraction of the first $r$-process peak with no heavy element nucleosynthesis. 
These scenarios add to the growing body of evidence against the contemporary viewpoint that a red kilonova is a ``smoking gun'' of lanthanide production.

In their respective analyses, both \cite{Rastinejad2022} and \cite{Yang24} infer ejecta properties using semi-analytic prescriptions which do not account for microphysical details such as photon reprocessing and radiative heating.
Inference using models without these microphysical details can lead to, among other things, overestimation of ejecta mass to compensate for the missing radiative heating contribution.
Our study encompasses these important effects and nevertheless recovers large ejecta masses, lending further confidence to our analysis favoring a collapsar nucleosynthesis scenario over one involving a merger.
Our study is partially motivated by the importance of including photon reprocessing and radiative heating microphysical effects when studying transients and understanding their discriminatory value in identifying transient progenitor candidates.
Such details can have significant downstream consequences, such as when inferring GRB progenitor populations and their effects on galactic chemical evolution \citep{Rastinejad25}.

In Section \ref{sec:inference}, we specify that the GRB230307A posteriors exhibit weak support for a high-velocity ejecta component.
In our fiducial simulation library, a comparison of weak $r$-process simulations with fixed $m_{\rm{ej}} = 0.1 M_\odot$ for $v_{\rm{ej}}/c = [0.05, 0.07, 0.10, 0.15, 0.18, 0.20, 0.30]$ finds greater variation across all broadband filters at early times ($t < 1$ day) compared to late times ($t > 1$ day).
This greater variation establishes that earlier observations carry more discriminating power for ejecta velocity; the relatively late-time ($t \gtrsim 1.5$ days) observations of GRB230307A thus set an upper limit on our ability to constrain the ejecta velocity, as evidenced by the weakly-supported high-velocity posterior.
We also recover significantly larger systematic uncertainty in our inference of GRB230307A, again owing largely to the limited dataset which lacks early-time observations and thus inhibits constraint on the full evolution of the light curves.
Our GRB230307A inference results highlight the community's need for established plans for rapid kilonova and gravitational-wave counterpart follow-up observations \citep{Magee21, TDAMM25, Desai25, Plestkova25, Singer25}.

We note that in this initial study, the interplay between nucleosynthesis and the detailed hydrodynamic evolution of the jet head are decoupled. Future studies incorporating a consistent treatment of the coupled photo-nuclear reaction network and the hydrodynamic mass flux through the jet head will be crucial to understanding the detailed $Y_{\rm{e}}$ evolution over the cocoon's dynamical time. These simulations could, in principle, also provide insight into the collapsar's supernova emission, or lackthereof as in the case of the GRBs discussed in this work and those in previous studies of long-GRBs which have shown similar non-detections of associated supernovae \citep[e.g.][]{Gendre13, Gupta25}.

\section{Conclusion}
\label{sec:conclusion}

Our work provides support for the idea that collapsars are a viable origin for the kilonovae associated with long-duration gamma-ray bursts and motivates additional detailed studies of photo-nuclear interactions in collapsar jet heads. 
We have demonstrated that the composition of weak $r$-process elements (up to proton number $Z \sim 50$ or mass number $A \sim 130$) can produce a spectrally evolving red component in kilonova emission. 
Thus a red kilonova does not inherently imply the synthesis of heavy $r$-process material --- a view that challenges contemporary interpretations. 
These insights prompt a reassessment of kilonova modeling assumptions and the inferred nucleosynthetic yields from such events. 
Future multi-messenger observations that include gravitational wave detections will be critical in distinguishing between the proposed mechanisms responsible for kilonovae associated with long-duration gamma-ray bursts.

\section{Acknowledgments}

Los Alamos National Laboratory (LANL) is operated by Triad National Security, LLC, for the National Nuclear Security Administration of U.S. Department of Energy (Contract No. 89233218CNA000001). 
M.~R. acknowledges support from the Information Science \& Technology Institute (ISTI) at LANL as an ISTI Postdoctoral Fellow and acknowledges PECASE award funds.
B.~L.~B. acknowledges support from the Advanced Simulation and Computing Program of LANL as a Metropolis Postdoc Fellow.
N.L.-R. acknowledges support from the Laboratory Directed Research and Development program of LANL project numbers 20230115ER and 20230217ER.
O.~K. is supported by the Laboratory Directed Research project 20240170ER.
S.~C., A~.G. and M.~R.~M. acknowledge support from the Laboratory Directed Research and Development program of LANL project numbers 20230052ER and 20240004DR. 
This research used resources provided by LANL through the Institutional Computing program.
This document has been assigned LA-UR-25-28887. 

\appendix

\section{Filtering Neutrons into the Cocoon}

\begin{figure}
    \centering
    \includegraphics[width=0.9\textwidth, trim={2.7cm 2cm 0 0}, clip]{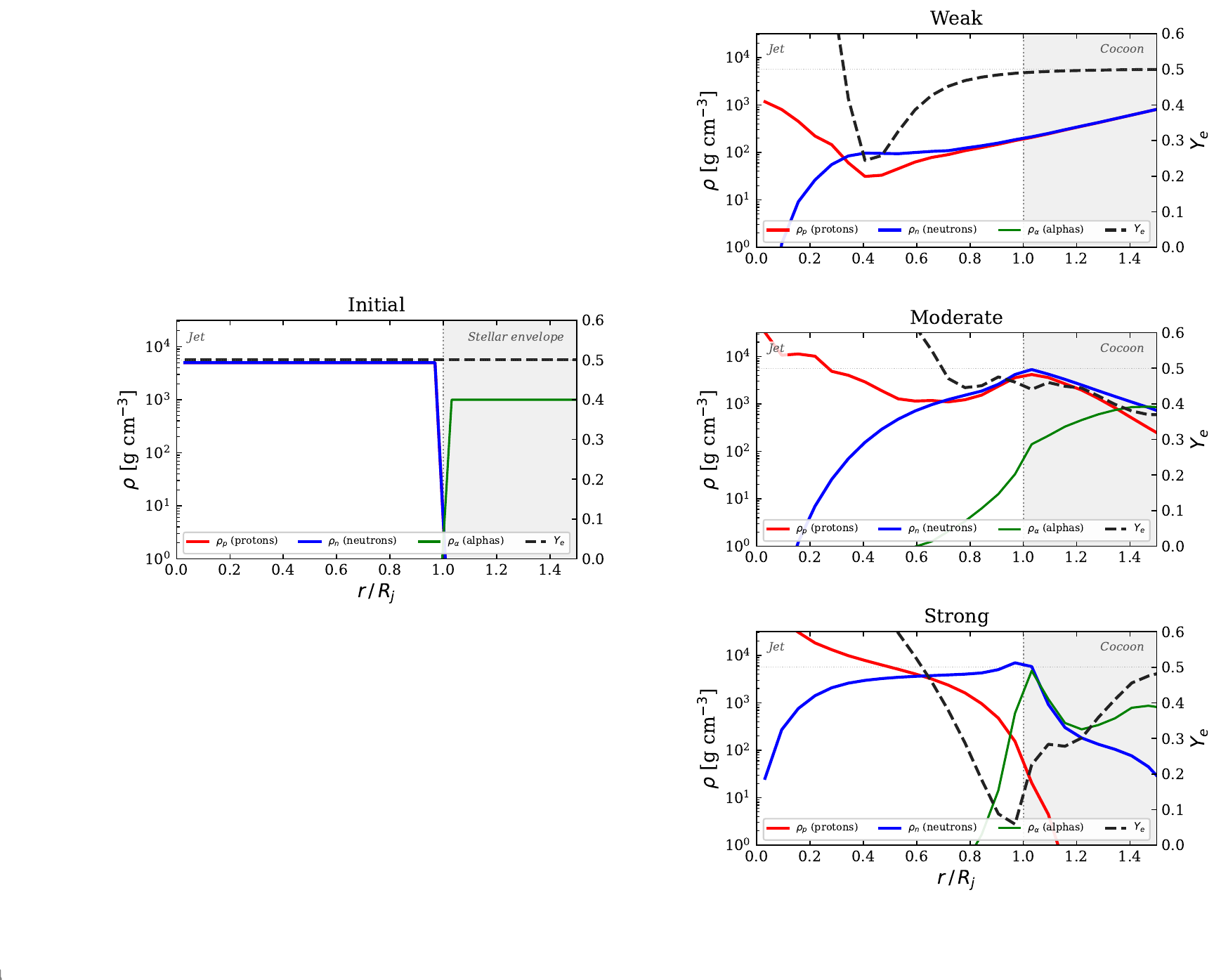}
    \caption{Density profiles and electron fraction $Y_e$ (dashed) from the three-fluid 2D MHD simulation. The left panel shows the initial conditions, with equal proton (red) and neutron (blue) densities in the jet core ($r < R_j$) and alpha particles (green) in the stellar envelope and developing cocoon ($r > R_j$). The three right panels show the evolved state for weak, moderate, and strong magnetic sieve regimes (increasing field strength from top to bottom). Each snapshot on the right panel is at different times in the range of $10^{-4}$ to $10^{-2}$ seconds.}
    \label{fig:sieve}
\end{figure}

The preferential filtering of neutrons into the cocoon while other charged particles (protons, alpha particles) remain trapped in the jet is described by a magnetic ``sieve" operating in the jet head. Here we describe the physical setup of this sieve and present a two-dimensional cylindrical magnetohydrodynamic (MHD) simulation showcasing its effects on particle separation. 

We assume that a toroidal magnetic field $B_\phi$ generated by the remnant compact object and accretion disk system propagates alongside the jet in a pinch-like configuration. The charged particles flowing in the jet create a current $J$, and the Lorentz force $F = J \times B$ felt by the charged particles points inward in $r$ direction, specifically toward $r=0$, the jet core. The sieve operates when the magnetic field dominates thermal pressure and the number of gyro-orbits completed by a proton between inter-particle collisions is large, which are satisfied at representative conditions in the jet head ($B \sim 10^{12}$~G, $T \sim 1$~GK, $\rho \sim 10^4$~g~cm$^{-3}$). 

To examine this effect, we perform a three-fluid (protons, neutrons, alpha particles) two-dimensional cylindrical MHD simulation. The simulation evolves the continuity, momentum, and energy equations for each species on an ($r$,$\phi$) grid. The simulation domain represents a cross-sectional slice through the jet head at fixed height $z$, focusing on the radial dynamics that drive species separation. Several initial magnetic field configurations were tested. In what follows, the initial magnetic field behaves as a Bennett z-pinch profile, $B_\phi(r) \propto r/(r^2 + R_j^2)$, where $R_j$ is the jet core radius. The jet interior is initialized with equal proton and neutron densities ($Y_e = 0.5$); the surrounding stellar envelope is composed of alpha particles in nuclear statistical equilibrium (NSE). The charged fluid is evolved with the CGL (Chew--Goldberger--Low) double-adiabatic closure, tracking perpendicular and parallel pressures independently, with isotropization proceeding through both wave--particle scattering and collisional relaxation channels. The magnetic field components $B_r$ and $B_\phi$ are evolved via the induction equation, including ambipolar diffusion. Transport is anisotropic: cross-field diffusion is suppressed by the same magnetization factor that governs the drag, while parallel transport follows along field lines. The resulting stiff system is solved with an implicit-explicit (IMEX) method.

Figure~\ref{fig:sieve} shows the results of our simulation for three realizations assuming a weak ($B \sim 10^{11}$ G), moderate $B \sim 5 \times 10^{11}$ G, and strong $B \gtrsim 10^{12}$ G magnetic field. In the left panel, we plot the $\theta$-averaged density of the three fluids (protons, neutrons, and alpha particles) as a function of the distance from the jet core in fractions of the jet radius $R_j$, noting that proton and neutron densities are equal ($Y_e = 0.5$) as stated earlier. In the right panels, we see the evolution of the individual fluids in the presence of the three different magnetic field realizations as well as the $Y_e$ evolution as a function of distance from the jet core. Notably, we find that the $Y_e$ near the jet-cocoon interface $r \sim R_j$ deviates from equilibrium in the moderate and strong magnetic field cases. In the strong case, we find that $Y_e \sim 0.3$ for $0.8 \leq r/R_j \leq 1.2$, suggesting that a relatively wide region around the jet-cocoon interface supports the required conditions for  $r$-process nucleosynthesis.

\bibliographystyle{aasjournal}
\bibliography{refs}

\end{document}